\documentclass[preprint]{aastex}
\usepackage{graphicx}

\newcommand{\teff}{\mbox{$T_{\rm eff}$}}
\newcommand{\logg}{\mbox{$\log g$}}
\newcommand{\vsini}{\mbox{$v \sin i_{\star}$}}
\newcommand{\mictrb}{\mbox{$\xi_{\rm t}$}}
\newcommand{\mactrb}{\mbox{$v_{\rm mac}$}}

\newcommand{\kms}{\mbox{km\,s$^{-1}$}}

\newcommand{\halpha}{\mbox{$H_\alpha$}}
\newcommand{\hbeta}{\mbox{$H_\beta$}}

\newcommand{\Msolar}{\mbox{$M_{\sun}$}}
\newcommand{\stara}{\mbox{WASP-77\,A}}
\newcommand{\starb}{\mbox{WASP-77\,B}}
\newcommand{\planet}{\mbox{WASP-77\,Ab}}

\shorttitle{\planet}
\shortauthors{Maxted et~al.}

\begin{document} 
\title {\planet: A transiting hot Jupiter planet in a wide binary
system.$^{\star}$}

\author{
P.~F.~L. Maxted\altaffilmark{1}, 
D.~R.~Anderson\altaffilmark{1}, 
A.~Collier Cameron\altaffilmark{2}, 
A.~P.~Doyle\altaffilmark{1}, 
A.~Fumel\altaffilmark{3},
M.~Gillon\altaffilmark{3}, 
C.~Hellier\altaffilmark{1}, 
E.~Jehin\altaffilmark{3},
M.~Lendl\altaffilmark{4},
F.~Pepe\altaffilmark{4}, 
D.~L.~Pollacco\altaffilmark{5,6}, 
D.~Queloz\altaffilmark{4}, 
D.~S\'egransan\altaffilmark{4}, 
B.~Smalley\altaffilmark{1}, 
J.~K.~Southworth \altaffilmark{1},
A.~M.~S.~Smith\altaffilmark{1}, 
A.~H.~M.~J.~Triaud\altaffilmark{4}, 
S.~Udry\altaffilmark{4},
R.~G. West\altaffilmark{7},
}

\altaffiltext{1}{Astrophysics Group, Keele University, Staffordshire, ST5 
 5BG, UK}
\altaffiltext{2}{SUPA, School of Physics and Astronomy, University of St.\
Andrews, North Haugh,  Fife, KY16 9SS, UK}
\altaffiltext{3}{Institut d'Astrophysique et de G\'eophysique, Universit\'e
de Li\`ege, All\'ee du 6 Ao\^ut, 17, Bat. B5C, Li\`ege 1, Belgium} 
\altaffiltext{4}{Observatoire astronomique de l'Universit\'e de Gen\`eve
51 ch. des Maillettes, 1290 Sauverny, Switzerland}
\altaffiltext{5}{Department of Physics, University of Warwick, Coventry, CV4
7AL}
\altaffiltext{6}{Astrophysics Research Centre, School of Mathematics \&
Physics, Queen's University, University Road, Belfast, BT7 1NN, UK}
\altaffiltext{7}{Department of Physics and Astronomy, University of 
Leicester, Leicester, LE1 7RH, UK}
\altaffiltext{$\star$}{Based on observations made with ESO Telescopes at the La
Silla Paranal Observatory under programme ID 088.C-0011.}

\begin{abstract}
 We report the discovery of a transiting planet with an orbital period of
1.36\,d orbiting the brighter component of the visual binary star
BD\,$-07^{\circ}\,436$. The host star, \stara, is a moderately bright G8\,V star
(V=10.3) with a metallicity close to solar ([Fe/H]$=0.0\pm0.1$). The companion
star, \starb, is a K-dwarf approximately 2 magnitudes fainter
at a separation of approximately 3\arcsec. The spectrum of \stara\ shows
emission in the cores of the Ca\,II H and K lines indicative of moderate
chromospheric activity. The WASP lightcurves show photometric variability with
a period of 15.3\,days and an amplitude of about 0.3\%  that is probably due
to the magnetic activity of the host star. We use an analysis of the
combined photometric and spectroscopic data to derive the mass and radius of
the planet ($1.76 \pm0.06M_{\rm Jup}$, $1.21 \pm0.02R_{\rm Jup}$). The age
of \stara\ estimated from its rotation rate ($\sim 1$\,Gyr) agrees with the
age estimated in a similar way for \starb\ ($\sim 0.6$\,Gyr)
but is in poor agreement with the age inferred by comparing its effective
temperature and density to stellar models ($\sim 8$\,Gyr). Follow-up
observations of \planet\ will make a useful contribution to our
understanding of the influence of  binarity and host star activity on the
properties of hot Jupiters.  \end{abstract}


\section{Introduction}

 Ground-based wide-angle surveys such as WASP \citep{2006PASP..118.1407P} and
HATnet \citep{2004PASP..116..266B} have now discovered more than 100
transiting hot Jupiter exoplanets around moderately bright stars ($8.5\la \rm
V\la 12.5$). The occurrence rate for hot Jupiters around solar-type stars is
known to be about 1\% from radial velocity surveys of bright stars
\citep{2012ApJ...753..160W}. The transit probability for a typical  hot
Jupiter with an orbital period $\approx 3$ days orbiting a solar-type star is
$\approx 10$\%. There are at least 340\,000 single FGK-type dwarf stars bright
enough to be accessible to survey such as WASP and HATnet
\citep{2006ApJ...638.1004A}, so many more transiting hot Jupiters remain to be
discovered. Increasing the sample of well-characterised transiting hot
Jupiters will clarify relations that may exist between parameters such as
period, mass, radius, etc., and so enable us to better understand the
formation, evolution and destruction of hot Jupiters \citep[e.g.,][]{	
2010ApJ...725.1995M, 2010ApJ...720.1569K,
2009MNRAS.396.1012D,2011ApJ...738....1B}. Finding new transiting
hot Jupiters will also reveal systems that have extreme properties or unusual
configurations that enable them to be characterised in ways not possible for
typical hot Jupiters, e.g.,  low density, bright planets such as WASP-17 that
can be characterised by transmission spectroscopy \citep{2011MNRAS.412.2376W}.
Detailed characterisation of a hot Jupiter using ground based observations is
always challenging given the small size of the signal due to the secondary
eclipse ($\la 0.1$\%) or the variation of transit depth with wavelength ($\la
0.01$\%). Such observations are made easier by the availability of a nearby
comparison star that can be used as a comparison source, particularly if the
companion is close enough to be included in the same entrance slit for
spectroscopic observations \citep[e.g.,][]{2012arXiv1208.4982S}.

 Here we report the discovery by the WASP survey  of a companion to the
brighter component of the visual binary star BD\,$-07^{\circ}~436$ with a mass
$\approx 1.7M_{\rm Jup}$ and a radius $\approx 1.2R_{\rm Jup}$. We find that
this star (\stara) is a G8\,V star showing moderate chromospheric activity.
The planet, \planet, is a typical hot Jupiter planet with an orbital period of
1.36 days. The companion star, \starb, is a K-dwarf approximately 2 magnitudes
fainter then \stara\ at a separation of approximately 3\arcsec. We show that
this star is physically associated with the star-planet system WASP-77A +
WASP-77Ab.

\section{Observations}
The WASP survey is described in \citet{2006PASP..118.1407P} and
\citet{2008ApJ...675L.113W} while a discussion of our candidate selection
methods can be found in \citet{2007MNRAS.380.1230C},
\citet{2008MNRAS.385.1576P}, and references therein. 

\begin{figure}
\plotone{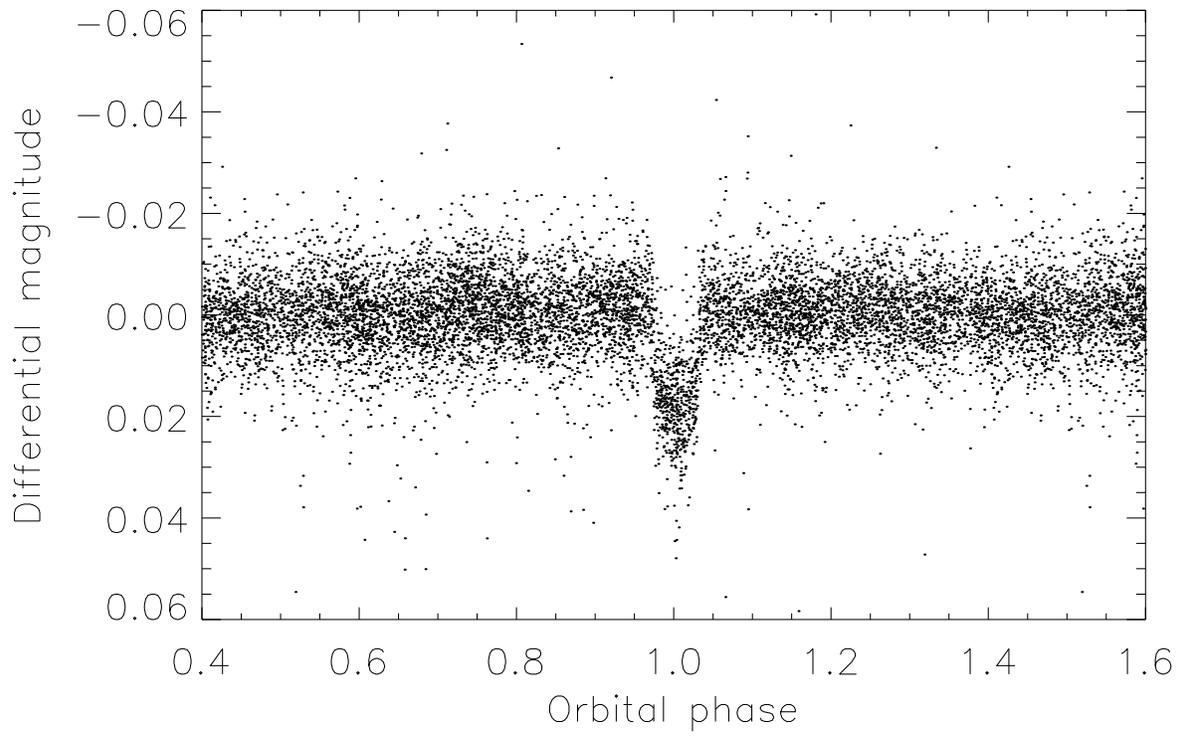} 
\caption{WASP photometry of WASP-77 plotted as a function of phase for a
period of 1.36003 days.
\label{wasplc} }
\end{figure}

 The star BD\,$-07^{\circ}\,436$ (WASP-77, 1SWASP J022837.22$-$070338.4) was
observed 5594 times by one camera on the WASP-South instrument from 2008 July
30 to 2008 December 12. The synthetic aperture radius used to measure the
flux of BD\,$-07^{\circ}\,436$ (48\arcsec) includes the flux from both
components of this visual binary star. We selected this star for follow-up
observations based on the characteristics  of the periodic transit-like
features detected in these data using the de-trending and transit detection
methods described in \cite{2007MNRAS.380.1230C}. The transits are also
detected with the same period  from 3316 observations obtained with the same
camera from 2009 August 2 to 2009 December 12. We have also analysed 898
observations obtained with a different camera obtained from 2008 August 18 to
2008 December 25. The WASP photometry is shown as a function of phase for a
period of 1.36003 days in Fig.~\ref{wasplc}.

 We obtained 11 radial velocity measurements  for \stara\ using the fibre-fed
CORALIE spectrograph on the Euler 1.2-m telescope located at La Silla, Chile.
Details of the instrument and data reduction can be found in
\cite{2000A&A...354...99Q} and \cite{2008ApJ...675L.113W}. The RV measurements
were performed using cross-correlation against a numerical mask generated from
a G2-type star and are given in Table~\ref{rv-data}, where we also provide the
bisector span, BS, which measures the asymmetry of the cross-correlation
function. The standard error of the the bisector span measurements is
estimated to be $2\sigma_{\rm RV}$. The amplitude and phase of the radial
velocity variations and the lack of any significant variation in the bisector
span from these data are consistent with the hypothesis that the transit
signal in the WASP photometry is due to a planetary mass object. However, the
diameter of the entrance aperture to the CORALIE spectrograph (2\arcsec) is
not quite sufficient to completely exclude light from the fainter component
contaminating the spectrum of the brighter component and {\it vice versa}, so
the CORALIE data by themselves are not sufficient to exclude the possibility
that the transit signal originates from \starb. We also obtained 4 radial
velocity measurements of \starb\ with CORALIE, but these are not reported here
because the spectra are clearly affected by contamination from the brighter
component. 

 Confirmation that the transit signal originates from the brighter component
of BD\,$-07^{\circ}\,436$ was obtained using the  60\,cm TRAPPIST telescope
\citep{2011Msngr.145....2J, 2012A&A...542A...4G} located at ESO La Silla
Observatory (Chile).  We obtained a sequence of 671 images of
BD\,$-07^{\circ}\,436$ covering the egress of a transit in good seeing
conditions using a z$^{\prime}$ filter on the night 2011 November 02. From a
selection of these images obtained in the best seeing we estimate that the
fainter component is  $3.22\pm0.05$ times fainter than the brighter component.
If the fainter star were responsible for the transit signal in the WASP
photometry then the eclipse in the lightcurve of this star would have a depth
of about 7\%. The stars are not completely resolved in these images, but
lightcurves of the two stars obtained using a synthetic aperture with a radius
of 3 pixels show a clear transit signal on the brighter component while the
fainter component is constant to within 2\%. This excludes the possibility
that the transit seen in the WASP photometry is due to a deep eclipse in the
lightcurve of the fainter component of the visual binary. We observed 3
further transits of BD\,$-07^{\circ}\,436$ using TRAPPIST on the nights 2011
November 01 (544 images), 05 November 2011 (1079 images) and 2011 December 01
(925 images). We also obtain a V-band lightcurve on the night 01 November 2011
(327 images) using the EulerCAM instrument on the Euler 1.2-m telescope
\citep{2012A&A...544A..72L}.  The flux ratio of the binary in the V-band
measured from these images is $5.00\pm0.13$.

\begin{figure} 
\plotone{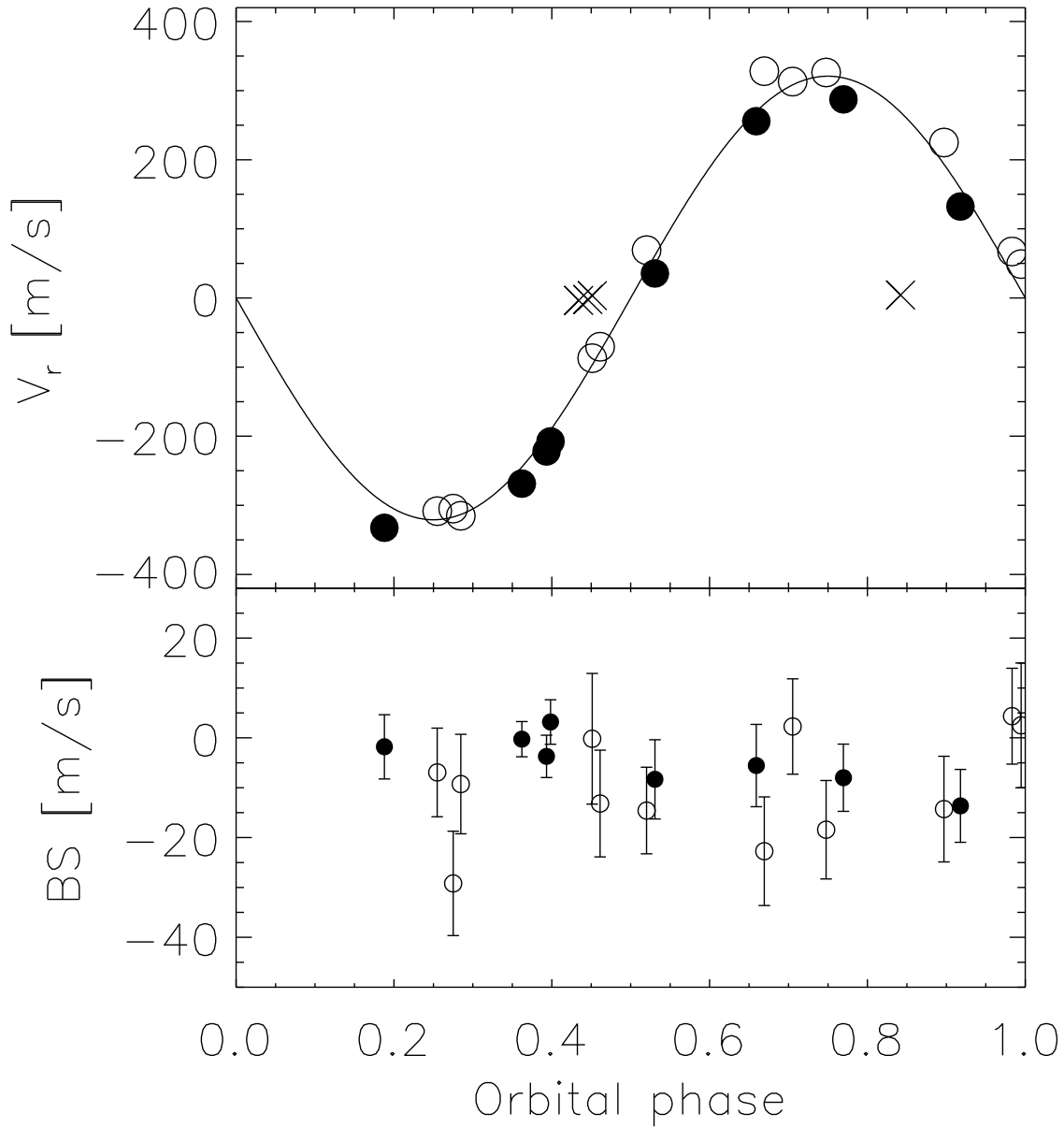} 
\caption{ Upper panel: Radial velocities of \stara\ relative to the
centre-of-mass velocity measured using HARPS (filled symbols) and CORALIE
(open symbols). Also shown are radial velocities of \starb\ relative to their
weighted mean value (crosses) and the best-fit circular orbit (solid line) for
\stara.  Lower panel: Lower panel: bisector span measurements for \stara\
(symbols as in upper panel).  
\label{rvphase} }
\end{figure}

 We used 8 spectra of \stara\ obtained with the HARPS spectrograph on the ESO
3.6m telescope (ESO programme ID 088.C-0011) to confirm that the radial velocity
signal seen in our CORALIE spectra originates from this star and not the
fainter component. The entrance aperture to HARPS has a diameter of 1\arcsec\
and the spectra were all obtained in good seeing so there is negligible
contamination of these spectra by light from the fainter component. Radial
velocities measured from these spectra using the same method as for our
CORALIE spectra are reported in Table~\ref{rv-data}. Also reported in this
table are 4  radial velocities for \starb\ measured in the same way. The
radial velocity of \starb\ is constant to within 10\,m\,s$^{-1}$. The radial
velocities of both stars and the bisector span measurements for \stara\ are
shown as a function of the transit phase in Fig.~\ref{rvphase}. 

 We obtained a series of 15,000 images of WASP-77 with an exposure time of 40
milliseconds using the 3-channel photometer ULTRACAM
\citep{2007MNRAS.378..825D} mounted on the 4.2-m William Herschel Telescope on
the night 2012 September 10. The pixel scale is approximately 0.30 arcseconds
per pixel and we used the u', g' and r' filters. We then selected  1\% of the
images in each channel with the best seeing  and combined them into three
high-resolution images, one for  each channel. We used these images to measure
the following magnitude difference between the components of the binary:
$\Delta$u' = $2.961\pm0.015$; $\Delta$g' = $2.156 \pm 0.004$; $\Delta$r' =
$1.701\pm0.007$. The separation of the components is 3.3 arcseconds. There are
no other stars visible in the small images we obtained so we do not have an
accurate astrometric solution for the images that we can use to estimate
robust errors on this value.

\section{Analysis}

\subsection{Stellar Parameters}

Eight individual HARPS spectra of \stara\  were co-added to produce a single
spectrum with an average S/N of around 80:1. Four co-added HARPS spectra of
\starb\ yielded a spectrum with an average S/N of 30:1. The standard pipeline
reduction products were used in the analysis. The analysis was performed using
the methods given in \citet{Doyle2012}. The \halpha\ and \hbeta\ lines were
used to give an initial estimate of the effective temperature (\teff). The
surface gravity (\logg) was determined from the Ca~{\sc i} lines at 6162{\AA}
and 6439{\AA} \citep{2010A&A...519A..51B}, along with the Na~{\sc i} D lines.
Additional \teff\ and \logg\ diagnostics were performed using the Fe lines. An
ionisation balance between Fe~{\sc i} and Fe~{\sc ii} was required, along with
a null dependence of the abundance on either equivalent width or excitation
potential \citep{2008A&A...478..487B}. This null dependence was also required
to determine the microturbulence (\mictrb). The parameters obtained from the
analysis are listed in Table~\ref{wasp77-params}. The elemental abundances
were determined from equivalent width measurements of several clean and
unblended lines, and additional least squares fitting of lines was performed
when required. The quoted error estimates include that given by the
uncertainties in \teff, \logg, and \mictrb, as well as the scatter due to
measurement and atomic data uncertainties. 

The projected stellar rotation velocity (\vsini, where $i_{\star}$ is the
inclination of the star's rotation axis) was determined by fitting the
profiles of several unblended Fe~{\sc i} lines. For \stara, a value for
macroturbulence (\mactrb) of 1.7 $\pm$ 0.3 {\kms} was assumed, based on the
calibration by \cite{2010MNRAS.405.1907B}. An instrumental FWHM of 0.04 $\pm$
0.01~{\AA} was determined from the resolution of the spectrograph. A best
fitting value of \vsini\ = 4.0 $\pm$ 0.2~\kms\ was obtained for \stara. For
\starb\ the macroturbulence was assumed to be zero, since for mid-K stars it
is expected to be lower than that of thermal broadening
\citep{2008oasp.book.....G}. A best fitting value of \vsini\ = 2.8 $\pm$
0.5~\kms\ was obtained for \starb.

There are emission peaks evident in the Ca~{\sc ii} H+K lines of \stara. The
signal-to-noise of the \starb\ spectra is too low to discern
any emission peaks.

\subsection{Rotation period}

\begin{figure} 
\begin{center}
\plotone{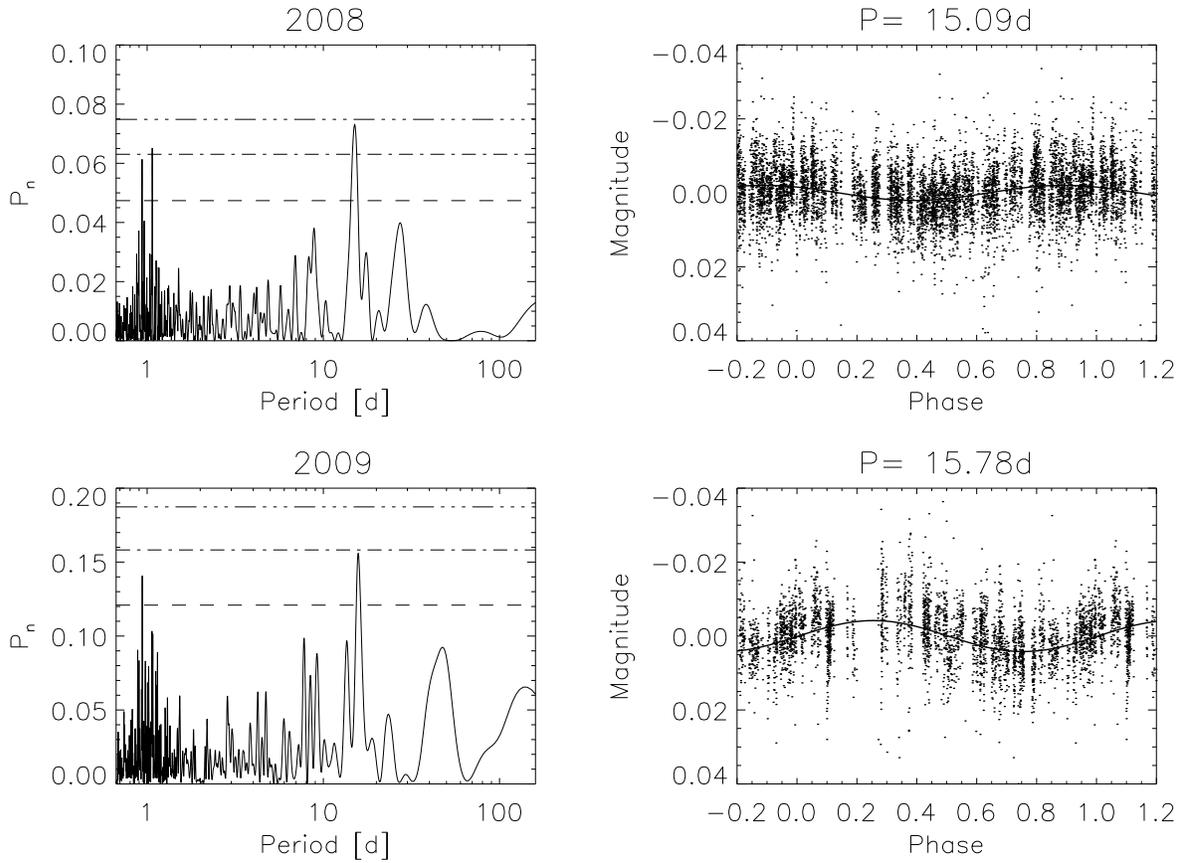}
\end{center}
\caption{{\it Left panels} Periodograms for the WASP data from
two different observing  seasons for WASP-77. Horizontal lines indicate false
alarm probability levels FAP=0.1,0.01,0.001. The year of
observation is noted in the title to each panel.
{\it Right panels} Lightcurves folded on the best period as noted in the
title.
\label{swlomb} }
\end{figure}

 The WASP lightcurves show a weak, periodic modulation with an amplitude of
about 0.3\,per~cent and a period of about 15\,days. This is likely to be a
signal of magnetic activity in \stara\ 
caused by star-spots modulating the apparent brightness as the star rotates.
WASP-77B may contribute to the variability of the
lightcurve on timescales of 10\,--\,20\,d, but it is unlikely to be the source
of the 15\,day modulation (see Section~\ref{discussion}).
We used the sine-wave fitting method described in \citet{2011PASP..123..547M}
to refine this estimate of the amplitude and period of the modulation.
Variability due to star spots is not expected to be coherent on long
timescales as a consequence of the finite lifetime of star-spots and
differential rotation in the photosphere so we analysed the two seasons of
data separately. We removed the transit signal from the data prior
to calculating the periodograms by subtracting a simple transit model from the
lightcurve. We calculated periodograms over 4096 uniformly spaced frequencies
from 0 to 1.5 cycles/day. The results for the two seasons of data are shown in
Fig.~\ref{swlomb}. The false alarm probability (FAP) levels shown in these
figures are calculated using a boot-strap Monte Carlo method also described in
\citet{2011PASP..123..547M}. There is a clear detection of a periodic
modulation with a period of 15.09 days in the 2008 data set (FAP=0.006). This
is confirmed by a detection at a period of 15.78 days, though with lower
significance, in the 2009 data set (FAP=0.052). We adopt a value of $P_{\rm
rot} = 15.4\pm 0.4$\,days for the period in the discussion below.

\subsection{Mass and radius of \planet}

To determine the planetary and orbital parameters, the HARPS radial velocity
measurements were combined with the photometry from TRAPPIST and EulerCAM in
a simultaneous least-squares fit using the Markov Chain Monte Carlo (MCMC)
technique. The details of this process are described in
\citet{2007MNRAS.380.1230C} and \citet{2008MNRAS.385.1576P}. Briefly, the
radial velocity data are modelled with a Keplerian orbit and the model of
\citet{2002ApJ...580L.171M} is used to fit the transits in the lightcurves. We
used the coefficients from \citet{2000A&A...363.1081C} for the
four-coefficient limb-darkening model.  We did not include the CORALIE radial
velocity data in the least-squares fit because it is unclear whether they are
affected by contamination from the companion star.  The TRAPPIST and EulerCAM
lightcurves were generated using a synthetic aperture radius large enough to
include the light from both stars so we applied a correction for the dilution
of the transit depth due to the light from the companion  prior to including
them in the MCMC analysis.

\begin{figure} 
\plotone{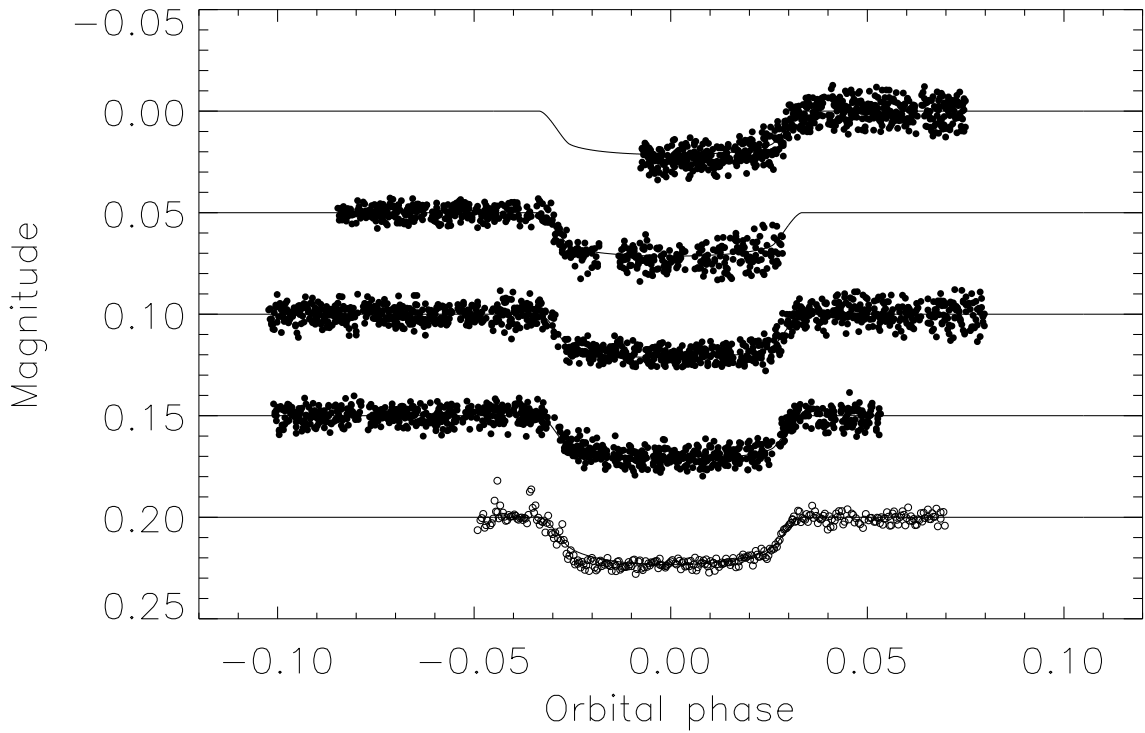} 
\caption{Photometry of \stara\ corrected for the dilution due to \starb. Data
from TRAPPIST are plotted with filled circles in date order from
top-to-botton. EulerCAM data are shown with open circles. The solid line shows
the lightcurve model for the parameters in Table~\ref{params}.
\label{lcfit} }
\end{figure}

 The baseline of the TRAPPIST and EulerCAM  observations is rather short, so
we used a measurement of the orbital period from an  analysis of the WASP
photometry as an additional constraint in the MCMC analysis  to the
TRAPPIST and EulerCAM data. The parameters of the model are given in
Table~\ref{params} and the model fits to the lightcurves are shown in
Fig.~\ref{lcfit}.  We have assumed that the orbit is circular because the
Lucy-Sweeney F-test applied to the results of a least-squares fit for an
eccentric orbit \citep{1971AJ.....76..544L} shows no evidence for a
non-circular  orbit ($p=0.18$, $e=0.008\pm0.005$). The parameters of the
transit model and the Keplerian orbit for the host star provide direct
estimates for the density of the star and the surface gravity of the planet.
To estimate the mass and radius of the planet we require an additional
constraint. In this case we derived the following relation specifically for
use with \stara\ and appropriate for stars with $0.8< M/\Msolar < 1.2$,
$-0.8<{\rm [Fe/H]}<0.3$ and $5000{\rm \,K}<\teff <6000{\rm \,K}$:
\[\log(M/\Msolar) = 0.0213 + 1.570 \log(\teff/5781{\rm \,K}) + 0.037
\log(\rho/\rho_{\sun}) + 0.097 {\rm [Fe/H]}. \] This equation is the result of
a least-squares fit the parameters of 19 stars in eclipsing binary systems
with accurately measured masses and radii.\footnote{\it
http://www.astro.keele.ac.uk/jkt/debcat/} The standard deviation of the
residuals from the fit is 0.051\Msolar. The standard error estimates for the
mass and radius of the star and planet given in Table~\ref{params} include
this contribution to the error budget.

We created two sets of EulerCAM and TRAPPIST lightcurves where the correction
for the dilution was increased or decreased by its standard error and
performed an MCMC analysis of these data in the same way as above. The
change in the system parameters between these two sets of data was used to
estimate the additional uncertainty on the system parameters due to the
uncertainty on the dilution factor. The additional uncertainty is found to be
small, e.g., the change in $(R_p/R_*)^2$ due to the uncertainty in the
dilution factor is 0.00007. This small additional uncertainty is included in
the standard errors given for all parameters affected in Table~\ref{params}.
  
\subsection{\label{discussion}Discussion}

\begin{figure} 
\begin{center}
\plotone{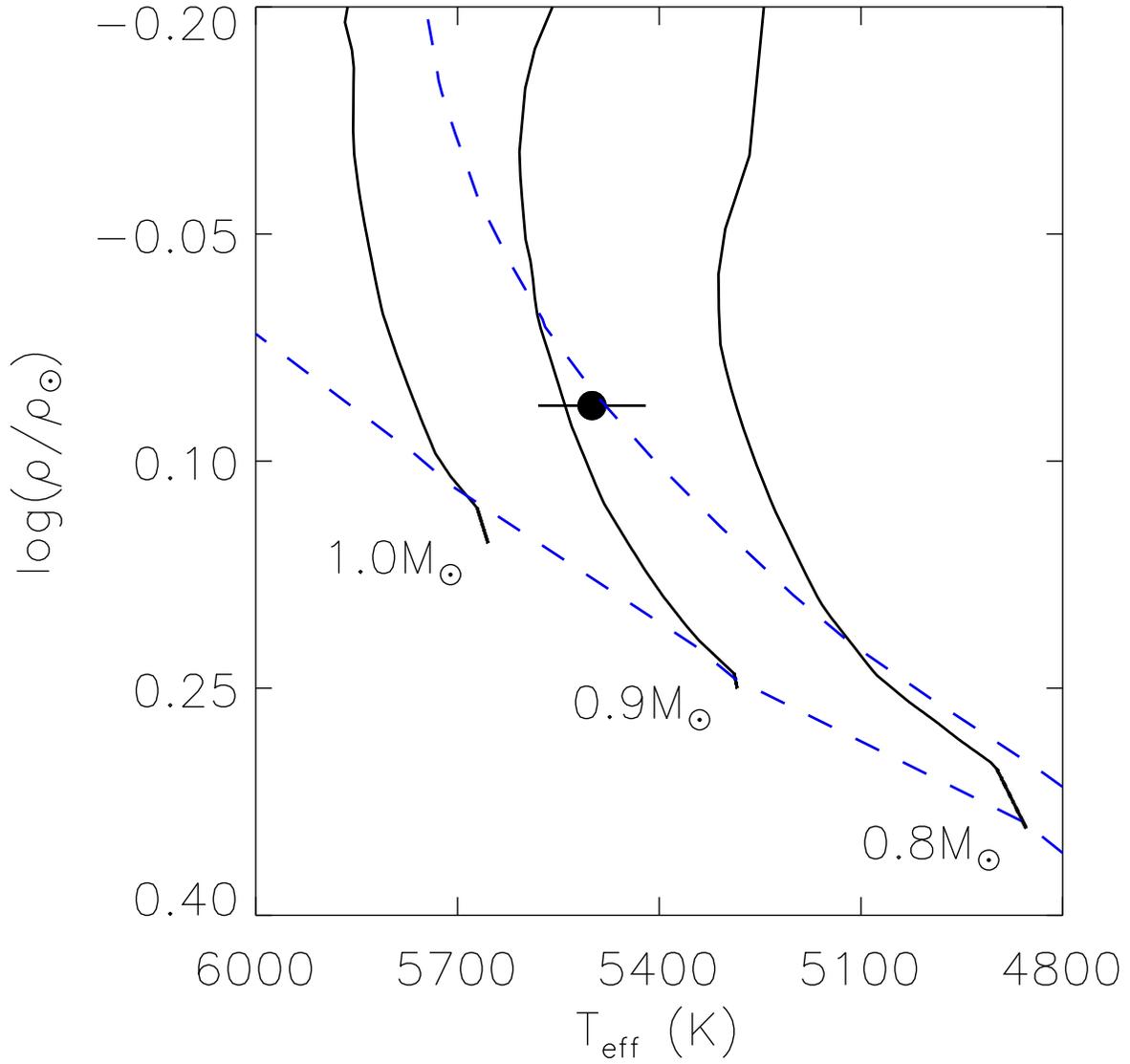}
\end{center}
\caption{Comparison of the effective temperature and stellar density of
\stara\ to the stellar models of  \citet{2000A&AS..141..371G}. Dashed lines
show isochrones for age of 10\,Myr and 10\,Gyr. Solid lines are evolutionary
tracks for stellar masses as indicated.
\label{trho} }
\end{figure}

 The density of \stara\ derived from our MCMC analysis is independent of any
assumption about the evolutionary state of the star, but our estimates for the
mass and radius for the  planet do assume that the stellar mass derived from
the empirical relation above is accurate. To test this assumption we also
compared the effective temperature and density of \stara\ to the stellar
models of \cite{2000A&AS..141..371G}.  The results are shown in
Fig.~\ref{trho}, where it can be seen the mass estimated from our MCMC
analysis ($M_* =  1.00  \pm  0.05$\Msolar) is consistent with these stellar
models, although the stellar models suggest a slightly lower mass ($\approx
0.89$\Msolar) and also suggest that the star is rather old ($\sim 8$\,Gyr),
i.e., slightly evolved. We discuss the reliability of these stellar models
further below. The mass and radius of \stara\ derived in our MCMC analysis
(Table~\ref{params}) agree well with the mass and radius expected for a  main
sequence star of the same spectral type (Table~\ref{wasp77-params}).

 The period of the modulation in the WASP lightcurve together with our
estimate for the radius of WASP-77\,A implies a rotation velocity $V_{\rm rot}
= 3.1 \pm 0.1$\,km\,s$^{-1}$ if we assume that this signal is due to the
rotation of this star. This compares with the spectroscopic estimate for the
projected rotation velocity $V_{\rm rot}\sin i_{\star} = 4.0\pm
0.2$\,km\,s$^{-1}$. The difference between these two values cannot be
explained by a mis-alignment between the star's rotation axis and the orbital
axis of \planet, but may be explained by an underestimate for the
macroturbulence (\mactrb) used in our analysis of the spectrum for \stara. The
calibration of \citet{2005ApJS..159..141V} suggests  a value of
\mactrb=3.5\kms. This additional line broadening  would reduce the value of
$V_{\rm rot}\sin i_{\star}$ estimated from the spectra sufficiently to make it
consistent with the hypothesis that the rotational signal in the WASP
lightcurves originates from \stara. The parameters listed in
Table~\ref{wasp77-params} are not affected by this uncertainty in the value of
\mactrb. The amplitude of the rotation signal in WASP lightcurves for K-type
stars with rotation periods $\sim 15$\,days can be as much a few percent
\citep{2009MNRAS.400..451C} so it also possible that the fainter component of
the visual binary contributes to the variability of the lightcurve, but it is
unlikely that the modulation of the lightcurve is due to the fainter component
alone.

 \stara\ and \starb\ appear to form a genuine physical binary
rather than a visual double star. The position angle (PA) and separation of
the stars estimated from the images we obtained in good seeing conditions with
EulerCAM and TRAPPIST and our high-resolution ULTRACAM images are consistent
with the values reported in The Washington Visual Double Star Catalog
\citep[WDS 02286$-$0704;][]{2001AJ....122.3466M}. That catalog reports that 7
observations were obtained between 1930 and 1933, during which time the
recorded separation varied from  2.9\arcsec\ to 3.2\arcsec. For comparison,
the proper motion  of \stara\ implies a change of position of 7.8\arcsec\
between 1930 and 2011.  The PA of the binary is $\approx 150^{\circ}$ whereas
the proper motion vector is approximately east-west, so the two components of
the binary clearly share a common proper motion.

 To estimate the distance to WASP-77, we used a least-squares fit to the  data
from \citet{2012ApJ...746..101B} to establish the following simple relation
between the angular diameter ($\theta_{\rm LD}$), H-band magnitude and
effective temperature of G-type dwarf stars. \[\theta_{\rm LD} = 3.206 -
0.679\log{\teff} -0.2 {\rm H} \] We then used this relation together with the
radius of \stara\ from Table~\ref{params} and the  apparent H-band magnitude
corrected for contribution from the fainter component to estimate a distance
$d=93\pm5$\,pc. We used the apparent V-band magnitude of WASP-77
\citep[$10.30\pm0.05$,][]{2000A&A...355L..27H} together with the V-band flux
ratio measured from our EulerCAM images to estimate an apparent V-band
magnitude for \starb\ of $12.05\pm0.06$, corresponding to an absolute
magnitude M$_{\rm V} = 7.2 \pm 0.1$, which is a typical value for a mid-K-type
dwarf star \citep{2008oasp.book.....G}. There is a small  difference between
the systemic velocity of \stara\ and the radial velocity of \starb\  ($\approx
1$\kms) but this is consistent with the orbital velocity expected for stars of
this mass and separation at the estimated distance of the binary. 

There is no significant detection of lithium in either \stara\ or
\starb. The equivalent width upper limits of 9~m\AA\ and
13~m\AA, correspond to an abundance upper limit of $\log A$(Li) $<$ 0.76 $\pm$
0.08 and $\log A$(Li) $<$ 0.10 $\pm$ 0.14 respectively. This implies an age at
least $\sim$0.5~Gyr \citep{2005A&A...442..615S} for both stars. 

 We can also estimate the ages of \stara\ and \starb\ based
on their rotation periods (gyrochronological age). The values of {\vsini} and
$R_{\star}$ for WASP-77A imply a rotation period $P_{\rm rot} = 14.2 \pm 1.7$
days (assuming $i_{\star}\approx 90$), which corresponds to a
gyrochronological age $\approx 1.0 ^{+0.5}_{-0.3}$~Gyr  using the
\citet{2007ApJ...669.1167B} relation. Similarly, an age $\approx
0.4^{+0.3}_{-0.2}$~Gyr is obtained for \starb\ from the rotation rate of
$P_{\rm rot} = 12.3 \pm 3.0$ days.  These age estimates are clearly
incompatible with the age for \stara\ we have derived using the stellar models
of \cite{2000A&AS..141..371G}. In this case, the agreement between the
gyrochronological ages of \stara\ and \starb\ suggest that gyrochronological age
estimates for hot Jupiter host stars are reliable, whereas ages estimated from
the effective temperature and density from isochrones may not be. We checked
that this is not a specific problem with this particular set of stellar models
by estimating the age of \stara\ using 5 additional sets of stellar models. The
results are shown in Table~\ref{stellarmodels} with stellar models noted as
follows: {\it Claret} \citep{2005A&A...440..647C}, $Y^2$
\citep{2004ApJS..155..667D}, {\it Teramo} \citep{2004ApJ...612..168P}, {\it
VRSS} \citep{2006ApJS..162..375V} and {\it DSEP} \citep{2008ApJS..178...89D}.
The models and method used are described in more detail in
\citet{2012MNRAS.426.1291S}. It can be seen that stellar models consistently
over-estimate the age of \stara\ compared to the gyrocronological age.  This
anomaly may well be related to the poor agreement between the observed radii of
some low mass stars and the radii predicted by stellar models, which in-turn is
thought to be a related to the rotation and/or magnetic activity in these stars
\citep{2011ApJ...728...48K, 2010ApJ...718..502M}. The mass of the star derived
from these stellar models is consistently lower than the value derived from our
empirical calibration, although not significantly different. The mass and radius
of the planet derived using these stellar models are consistent with the values
derived using our empirical calibration.

\section{Conclusions}

 The brighter component of the visual binary star BD\,$-07^{\circ}\,436$,
\stara, shows transits every 1.36 days caused by a  hot Jupiter companion,
\planet,  with a mass $\approx 1.8$M$_{\rm Jup}$ and a radius $\approx
1.2$R$_{\rm Jup}$. The radial velocities of the two components of the binary
reported here strengthen the conclusion that the two stars are physically
associated. The gyrochronological ages for the two stars agree and suggest an
age $\sim 0.5$\,--\,$1$\,Gyr for the stars. If the gyrochronological age for
\stara\ is correct, then this star has a lower density than predicted by
standard stellar models. This is a phenomenon seen in other stars of similar
mass and is likely to be related to the magnetic activity observed in this
star.

\acknowledgments
WASP-South is hosted by the South African Astronomical Observatory and we are
grateful for their ongoing support and assistance. Funding for WASP comes from
consortium universities and from the UK's Science and Technology Facilities
Council.  We thank the ULTRACAM team taking the observations of WASP-77
presented here. TRAPPIST is funded by the Belgian Fund for Scientific Research
(Fond National de la Recherche Scientifique, FNRS) under the grant FRFC
2.5.594.09.F, with the participation of the Swiss National Science Fundation
(SNF). M. Gillon and E. Jehin are FNRS Research Associates.

\bibliographystyle{pasp}
\bibliography{wasp}

\begin{table} 
\begin{center}
\caption{Radial velocity measurements for \stara\ and \starb. 
\label{rv-data} }
\begin{tabular}{lrrr} 
\tableline\tableline 
BJD&\multicolumn{1}{l}{RV} & 
\multicolumn{1}{l}{$\sigma_{\rm RV}$} & 
\multicolumn{1}{l}{BS}\\ 
$-$2\,450\,000  & (km s$^{-1}$) & (km s$^{-1}$) & (km s$^{-1}$)\\ 
\tableline
\multicolumn{4}{l}{\stara, CORALIE} \\
5069.7803 &  1.3246 & 0.0049 & $-$0.02828 \\
5163.6091 &  1.3362 & 0.0052 & $-$0.04828 \\
5169.6920 &  1.9673 & 0.0049 & $-$0.03717 \\
5170.6625 &  1.5706 & 0.0053 & $-$0.03203 \\
5188.6258 &  1.9696 & 0.0054 & $-$0.04100 \\
5856.7095 &  1.8655 & 0.0053 & $-$0.03387 \\
5914.6783 &  1.7105 & 0.0043 & $-$0.03349 \\
5915.6775 &  1.3328 & 0.0044 & $-$0.02568 \\
5916.6681 &  1.7087 & 0.0048 & $-$0.01432 \\
5917.6500 &  1.9544 & 0.0047 & $-$0.01638 \\
5918.6645 &  1.5542 & 0.0065 & $-$0.01908 \\
\noalign{\smallskip}
\multicolumn{4}{l}{\stara, HARPS} \\
5832.8615 & 1.4626 & 0.0020 & 0.0365 \\	
5832.9040 & 1.5111 & 0.0024 & 0.0352 \\	
5832.9110 & 1.5233 & 0.0025 & 0.0410 \\	
5889.7458 & 1.4063 & 0.0036 & 0.0307 \\	
5890.5370 & 2.0220 & 0.0037 & 0.0375 \\	
5890.7386 & 1.8656 & 0.0041 & 0.0180 \\	
5891.5721 & 1.7670 & 0.0044 & 0.0260 \\	
5891.7468 & 1.9885 & 0.0046 & 0.0432 \\	
\noalign{\smallskip}
\multicolumn{4}{l}{\starb, HARPS} \\
5832.8705 & 2.7508 & 0.0047 &  0.0168 \\	
5832.8885 & 2.7521 & 0.0061 &  0.0373 \\	
5832.8960 & 2.7577 & 0.0063 &  0.0172 \\	
5890.5450 & 2.7586 & 0.0057 &  0.0203 \\	
\noalign{\smallskip}
\tableline 
\end{tabular} 
\end{center}
\end{table}

\begin{table}
\begin{center}
\caption{Stellar parameters of \stara\ and \starb.
 Abundances are relative to the solar values obtained by
\citep{2009ARA&A..47..481A}.
\label{wasp77-params}
}
\begin{tabular}{lrr}
\tableline\tableline
Parameter  & \stara & \starb \\ \tableline
\teff\ [K]   &   5500 $\pm$ 80 & 4700 $\pm$ 100\\
\logg      &   4.33 $\pm$ 0.08 & 4.6 $\pm$ 0.15\\
\mictrb\ [\kms]   &   0.8 $\pm$ 0.1  & \\
\vsini\ [\kms]    &   4.0 $\pm$ 0.2 & 2.8 $\pm$ 0.5 \\
{[Fe/H]}   &   0.00 $\pm$ 0.11 &$-$0.12 $\pm$ 0.19\\
{[Mg/H]}   &   0.23 $\pm$ 0.04 &   0.09 $\pm$ 0.10\\
{[Ca/H]}   &$-$0.02 $\pm$ 0.13 &$-$0.01 $\pm$ 0.13\\
{[Sc/H]}   &$-$0.03 $\pm$ 0.07 &   0.14 $\pm$ 0.18\\
{[Ti/H]}   &$-$0.02 $\pm$ 0.10 &   0.15 $\pm$ 0.21\\
{[V/H]}	   &$-$0.03 $\pm$ 0.09 &   0.39 $\pm$ 0.15\\
{[Cr/H]}   &   0.00 $\pm$ 0.06 &   0.07 $\pm$ 0.22\\
{[Mn/H]}   &   0.14 $\pm$ 0.14 &   0.06 $\pm$ 0.34\\
{[Co/H]}   &$-$0.08 $\pm$ 0.10 &   0.24 $\pm$ 0.27\\
{[Ni/H]}   &$-$0.01 $\pm$ 0.08 &   0.00 $\pm$ 0.19\\
{[Y/H]}    &   0.04 $\pm$ 0.08 &\\
$\log A$(Li)  & $< 0.76 \pm$ 0.08 & $<0.10 \pm$ 0.14\\
Mass  [$M_{\sun}$]    &   1.00 $\pm$ 0.07 & 0.71 $\pm$ 0.06  \\
Radius [$R_{\sun}$]    &   1.12 $\pm$ 0.12 & 0.69 $\pm$ 0.12  \\
Sp. Type\tablenotemark{a}  &   G8 & K5\\
\tableline
\end{tabular}
\tablenotetext{a}{Spectral type estimated from \teff\
using the table in \cite{2008oasp.book.....G}.}
\end{center}
\end{table}

\begin{table*}
\begin{center}
\caption{System parameters and $1\sigma$ error limits derived from the MCMC
analysis. \label{params}}
\begin{tabular}{lcrl}
\tableline \tableline
Parameter & Symbol & Value & Units \\
 \tableline 
Orbital period & $P$ & $ 1.3600309 \pm 0.0000020 $ & days \\ 
Transit epoch & $T_0$ & $ 2455870.44977 \pm 0.00014 $ & BJD \\ 
Planet/star area ratio & $(R_p/R_*)^2$ & $  0.01693 \pm  0.00017 $ & \\ 
Transit duration & $t_T$ & $  0.09000 \pm  0.00035$ & days \\ 
\noalign{\smallskip}
Impact parameter & $b$ & $  0.06^{+ 0.07}_{-0.05}$ &  \\ 
\noalign{\smallskip}
Stellar reflex velocity & $K_1$ & $ 0.3219\pm0.0039$ & \mbox{km s$^{-1}$} \\ 
Centre-of-mass velocity & $\gamma$ & $ 1.6845 \pm 0.0004 $ & \mbox{km s$^{-1}$} \\
\noalign{\smallskip}
Orbital inclination & $i$ & $89.4^{+0.4}_{-0.7}$ & $^{\circ}$\\ 
\noalign{\smallskip}
Stellar density & $\rho_*$ & $ 1.157^{+0.016}_{-0.020}$ & $\rho_\sun$ \\ 
Stellar mass & $M_*$ & $ 1.002 \pm  0.045$ & $M_\sun$ \\ 
\noalign{\smallskip}
Stellar radius & $R_*$ & $ 0.955 \pm 0.015 $ & $R_\sun$ \\ 
\noalign{\smallskip}
Orbital semi-major axis & $a$ & $  0.0240  \pm0.00036$ & $AU$ \\ 
\noalign{\smallskip}
Planet mass & $M_p$ & $  1.76 \pm 0.06 $ & $M_{\rm Jup}$ \\ 
\noalign{\smallskip}
Planet radius & $R_p$ & $  1.21 \pm 0.02$ &$R_{\rm Jup}$\tablenotemark{a}\\ 
Planet surface gravity & $\log g_p$ & $3.441 \pm 0.008 $ & [cgs] \\ 
Planet density & $\rho_p$ & $1.00 \pm 0.03$ & $\rho_{\rm Jup}$ \\ 
\tableline
\end{tabular}
\tablenotetext{a}{$R_{\rm Jup}$ = 71492\,km}
\end{center}
\end{table*}

\begin{table} 
\begin{center}
\caption{\label{stellarmodels} Mass, radius and age
of \stara\ and \planet\ derived using different sets of stellar models.}
\begin{tabular}{lrrrrr}
\tableline \tableline
& \multicolumn{1}{l}{\it Claret} & \multicolumn{1}{l}{$Y^2$} & 
\multicolumn{1}{l}{\it Teramo}  & \multicolumn{1}{l}{\it VRSS}  &
\multicolumn{1}{l}{\it DSEP} \\
\tableline
$M_{\rm A}$ ($M_{\sun}$) &$ 0.948 \pm 0.055 $&$ 0.940 \pm 0.048 $&$ 0.889 \pm 0.050 $&$ 0.893 \pm 0.048 $&$ 0.900 \pm 0.008 $ \\
$R_{\rm A}$ ($R_{\sun}$) &$ 0.938 \pm 0.019 $&$ 0.935 \pm 0.017 $&$ 0.918 \pm 0.018 $&$ 0.919 \pm 0.017 $&$ 0.922 \pm 0.007 $ \\
$M_{\rm b}$ ($M_{\rm Jup}$) &$ 1.695 \pm 0.069 $&$ 1.686 \pm 0.061 $&$ 1.624 \pm 0.064 $&$ 1.629 \pm 0.062 $&$ 1.637 \pm 0.024 $ \\
$R_{\rm b}$ ($R_{\rm Jup}$) &$ 1.186 \pm 0.025 $&$ 1.183 \pm 0.022 $&$ 1.161 \pm 0.024 $&$ 1.163 \pm 0.023 $&$ 1.166 \pm 0.011 $ \\
Age (Gyr) &$ 6.8^{+4.7}_{-2.4}$&$5.8^{+2.9}_{-1.8}$&$10.4^{+1.6}_{-4.3}$ &$9.5^{+1.6}_{-4.4}$ &$ 8.2^{+0.9}_{-1.6}$ \\
\tableline 
\end{tabular}
\end{center}
\end{table}

\end{document}